\documentstyle[twoside,epsf,psfig,times]{article}
\def\beqa{\begin{eqnarray}}
\def\eeqa{\end{eqnarray}}
\def\beq{\begin{equation}}
\def\eeq{\end{equation}}
\catcode`\@=11
\long\def\@makefntext#1{
\protect\noindent \hbox to 3.2pt {\hskip-.9pt  
$^{{\eightrm\@thefnmark}}$\hfil}#1\hfill}               %CAN BE USED 

\def\@makefnmark{\hbox to 0pt{$^{\@thefnmark}$\hss}}    %ORIGINAL 

\def\ps@myheadings{\let\@mkboth\@gobbletwo
\def\@oddhead{\hbox{}
\rightmark\hfil\eightrm\thepage}   
\def\@oddfoot{}\def\@evenhead{\eightrm\thepage\hfil
\leftmark\hbox{}}\def\@evenfoot{}
\def\sectionmark##1{}\def\subsectionmark##1{}}

\oddsidemargin=\evensidemargin
\newcounter{sectionc}\newcounter{subsectionc}\newcounter{subsubsectionc}
\renewcommand{\section}[1] {\vspace{12pt}\addtocounter{sectionc}{1} 
\setcounter{subsectionc}{0}\setcounter{subsubsectionc}{0}\noindent 
        {\tenbf\thesectionc. #1}\par\vspace{5pt}}
\renewcommand{\subsection}[1] {\vspace{12pt}\addtocounter{subsectionc}{1} 
\setcounter{subsubsectionc}{0}\noindent 
{\bf\thesectionc.\thesubsectionc. {\kern1pt \bfit #1}}\par\vspace{5pt}}
\renewcommand{\subsubsection}[1] {\vspace{12pt}\addtocounter{subsubsectionc}{1}
        \noindent{\tenrm\thesectionc.\thesubsectionc.\thesubsubsectionc.
        {\kern1pt \tenit #1}}\par\vspace{5pt}}
\newcommand{\nonumsection}[1] {\vspace{12pt}\noindent{\tenbf #1}
        \par\vspace{5pt}}

\newcounter{appendixc}
\newcounter{subappendixc}[appendixc]
\newcounter{subsubappendixc}[subappendixc]
\renewcommand{\thesubappendixc}{\Alph{appendixc}.\arabic{subappendixc}}
\renewcommand{\thesubsubappendixc}
        {\Alph{appendixc}.\arabic{subappendixc}.\arabic{subsubappendixc}}

\renewcommand{\appendix}[1] {\vspace{12pt}
        \refstepcounter{appendixc}
        \setcounter{figure}{0}
        \setcounter{table}{0}
        \setcounter{lemma}{0}
        \setcounter{theorem}{0}
        \setcounter{corollary}{0}
        \setcounter{definition}{0}
        \setcounter{equation}{0}
        \renewcommand{\thefigure}{\Alph{appendixc}.\arabic{figure}}
        \renewcommand{\thetable}{\Alph{appendixc}.\arabic{table}}
        \renewcommand{\theappendixc}{\Alph{appendixc}}
        \renewcommand{\thelemma}{\Alph{appendixc}.\arabic{lemma}}
        \renewcommand{\thetheorem}{\Alph{appendixc}.\arabic{theorem}}
        \renewcommand{\thedefinition}{\Alph{appendixc}.\arabic{definition}}
        \renewcommand{\thecorollary}{\Alph{appendixc}.\arabic{corollary}}
        \renewcommand{\theequation}{\Alph{appendixc}.\arabic{equation}}
        \noindent{\tenbf Appendix \theappendixc #1}\par\vspace{5pt}}
\newcommand{\subappendix}[1] {\vspace{12pt}
        \refstepcounter{subappendixc}
        \noindent{\bf Appendix \thesubappendixc. {\kern1pt \bfit #1}}
        \par\vspace{5pt}}
\newcommand{\subsubappendix}[1] {\vspace{12pt}
        \refstepcounter{subsubappendixc}
        \noindent{\rm Appendix \thesubsubappendixc. {\kern1pt \tenit #1}}
        \par\vspace{5pt}}

\newcommand{\textlineskip}{\baselineskip=13pt}
\newcommand{\smalllineskip}{\baselineskip=10pt}

\newcommand{\copyrightheading}[1]
        {\vspace*{-2.5cm}\smalllineskip{\flushleft
        {\footnotesize International Journal of Modern Physics D, #1}\\
        {\footnotesize \copyright\kern2pt World Scientific Publishing
         Company}\\
         }}

\newcommand{\publisher}[2]{{\begin{center}\footnotesize\smalllineskip 
        Received #1\\
        Revised #2
        \end{center}
        }}

\def\abstracts#1#2#3{{
        \centering{\begin{minipage}{4.5in}\footnotesize\baselineskip=10pt
        \parindent=0pt #1\par 
        \parindent=15pt #2\par
        \parindent=15pt #3
        \end{minipage}}\par}} 

\renewenvironment{thebibliography}[1]
        {\frenchspacing
         \ninerm\baselineskip=11pt
         \begin{list}{\arabic{enumi}.}
        {\usecounter{enumi}\setlength{\parsep}{0pt}     
         \setlength{\leftmargin 12.7pt}{\rightmargin 0pt}%FOR 1--9 ITEMS
         \setlength{\itemsep}{0pt} \settowidth
        {\labelwidth}{#1.}\sloppy}}{\end{list}}

\newcounter{itemlistc}
\newcounter{romanlistc}
\newcounter{alphlistc}
\newcounter{arabiclistc}

\newcommand{\fcaption}[1]{
        \refstepcounter{figure}
        \setbox\@tempboxa = \hbox{\footnotesize Fig.~\thefigure. #1}
        \ifdim \wd\@tempboxa > 5in
           {\begin{center}
        \parbox{5in}{\footnotesize\smalllineskip Fig.~\thefigure. #1}
            \end{center}}
        \else
             {\begin{center}
             {\footnotesize Fig.~\thefigure. #1}
              \end{center}}
        \fi}

\newcommand{\tcaption}[1]{
        \refstepcounter{table}
        \setbox\@tempboxa = \hbox{\footnotesize Table~\thetable. #1}
        \ifdim \wd\@tempboxa > 5in
           {\begin{center}
        \parbox{5in}{\footnotesize\smalllineskip Table~\thetable. #1}
            \end{center}}
        \else
             {\begin{center}
             {\footnotesize Table~\thetable. #1}
              \end{center}}
        \fi}

\def\@citex[#1]#2{\if@filesw\immediate\write\@auxout
        {\string\citation{#2}}\fi
\def\@citea{}\@cite{\@for\@citeb:=#2\do
        {\@citea\def\@citea{,}\@ifundefined
        {b@\@citeb}{{\bf ?}\@warning
        {Citation `\@citeb' on page \thepage \space undefined}}
        {\csname b@\@citeb\endcsname}}}{#1}}

\newif\if@cghi
\def\cite{\@cghitrue\@ifnextchar [{\@tempswatrue
        \@citex}{\@tempswafalse\@citex[]}}
\def\citelow{\@cghifalse\@ifnextchar [{\@tempswatrue
        \@citex}{\@tempswafalse\@citex[]}}
\def\@cite#1#2{{$\null^{#1}$\if@tempswa\typeout
        {IJCGA warning: optional citation argument 
        ignored: `#2'} \fi}}

\def\pmb#1{\setbox0=\hbox{#1}
        \kern-.025em\copy0\kern-\wd0
        \kern.05em\copy0\kern-\wd0
        \kern-.025em\raise.0433em\box0}

\def\fnt#1#2{\footnotetext{\kern-.3em
        {$^{\mbox{\scriptsize #1}}$}{#2}}}

\def\fpage#1{\begingroup
\voffset=.3in
\thispagestyle{empty}\begin{table}[b]\centerline{\footnotesize #1}
        \end{table}\endgroup}

\def\runninghead#1#2{\pagestyle{myheadings}
\markboth{{\protect\footnotesize\it{\quad #1}}\hfill}
{\hfill{\protect\footnotesize\it{#2\quad}}}}
\font\tenrm=cmr10
\font\tenit=cmti10 
\font\tenbf=cmbx10
\font\bfit=cmbxti10 at 10pt
\font\ninerm=cmr9

\font\eightrm=cmr8

\def\qed{\hbox{${\vcenter{\vbox{                  %HOLLOW SQUARE
   \hrule height 0.4pt\hbox{\vrule width 0.4pt height 6pt
   \kern5pt\vrule width 0.4pt}\hrule height 0.4pt}}}$}}

\begin{document}
\setlength{\textheight}{7.7truein}    %FOR 2ND PAGE ONWARDS

\runninghead{Bulk Viscous Solutions to the Field Equations and Decelaration Parameter Revisited } {A.
 Pradhan and Iotemshi I.}

\normalsize\textlineskip
\thispagestyle{empty}
\setcounter{page}{1}

\copyrightheading{}             {Vol.~0, No.~0 (2002) 000--000}

\vspace*{0.88truein}

\fpage{1}

\centerline{\bf BULK VISCOUS SOLUTIONS TO THE FIELD EQUATIONS}
\vspace*{0.035truein}
\centerline{\bf AND THE DECELERATION PARAMETER-REVISITED}  
\vspace*{0.37truein}
%\centerline{}
%\vspace*{0.015truein}
%\centerline{}
%\baselineskip=10pt
%\centerline{}
%\vspace*{10pt}
\centerline{\footnotesize ANIRUDH PRADHAN\footnote{E-mail: acpradhan@yahoo.com,
pradhan@iucaa.ernet.in (Corresponding Author)},~~ IOTEMSHI I.}
\vspace*{0.015truein}
\centerline{\footnotesize\it Department of Mathematics, Hindu Post-graduate College,}
\baselineskip=10pt
\centerline{\footnotesize\it Zamania, Ghazipur 232 331, India}
\vspace*{10pt}
%\centerline{}
%\vspace*{0.015truein}
%\centerline{\footnotesize}
%\baselineskip=10pt
%\centerline{\footnotesize}
\vspace*{0.225truein}
\publisher{(received date)}{(revised date)}

\vspace*{0.21truein}
\abstracts{We utilise a form for the Hubble parameter to generate a number of solutions 
to the Einstein field equations with variable cosmological constant and 
variable gravitational constant in the presence of a bulk viscous fluid. The
Hubble law utilised yields a constant value for the deceleration parameter. A
new class of solutions is presented in the Robertson-Walker spacetimes. The
coefficient of bulk viscosity is assumed to be a power function of the mass
density. For a class of solutions, the deceleration parameter is negative
which is consistent with the supernovae Ia observations.}{}{}

%\vspace*{10pt}
%\keywords{The contents of the keywords}

%\textlineskip                  %) USE THIS MEASUREMENT WHEN THERE IS
%\vspace*{12pt}                 %) NO SECTION HEADING

\vspace*{1pt}\textlineskip      %) USE THIS MEASUREMENT WHEN THERE IS
\section{Introduction}
\vspace*{-0.5pt}
\noindent
Einstein proposed his General Theory of Relativity as a geometric theory.
He introduced a cosmological constant into his field equations in order 
to obtain a static cosmological model as without the cosmological term, the
field equations would admit only non-static solutions. The Einstein field
equations are a coupled system of highly nonlinear differential equations and
we seek physical solutions for applications in cosmology and astrophysics. In
order to solve the field equations we normally assume a form for the matter 
content or suppose that spacetime admits Killing vector symmetries \cite{ref1}.  
Solutions to the field equations may also be generated by applying a law of
variation for Hubble's parameter.
It is interesting to observe that this law yields a constant value for the 
deceleration parameter. The variation of
Hubble's law assumed is not inconsistent with observation and has the
advantage of providing simple functional forms of the scale factor. In the
simplest case the Hubble law yields a constant value for the deceleration
parameter. In earlier literature cosmological models with a constant deceleration
parameter have been studied by Berman \cite{ref2,ref3}, Berman and Gomide \cite{ref4},
Johri and Desikan \cite{ref5}, Singh and Desikan \cite{ref6}, Pradhan et al
\cite{ref7} and others. 
\newline
\par
Models with a relic cosmological constant $\Lambda$ have received considerable 
attention recently among researchers for various reasons 
(see Refs. {\cite{ref8}}$^-${\cite{ref12}} and references therein). Some of the 
recent discussions on the cosmological constant ``problem'' and on cosmology 
with a time-varying cosmological constant by Ratra and Peebles \cite{ref13}, 
Dolgov \cite{ref14}$^-$\cite{ref16} and Sahni and Starobinsky \cite{ref17}
point out that in the absence of any interaction with matter or radiation, the 
cosmological constant remains a ``constant'', however, in the presence of
interactions with matter or radiation, a solution of Einstein equations and the 
assumed equation of covariant conservation of stress-energy with a time-varying 
$\Lambda$ can be found. For these solutions, conservation of energy requires 
decrease in the energy density of the vacuum component to be compensated by a 
corresponding increase in the energy density of matter or radiation. Earlier 
researchers on this topic, are contained in Zeldovich \cite{ref18}, 
Weinberg \cite{ref9} and Carroll, Press and Turner \cite{ref19}. Recent
observations by Perlmutter {\it et al} \cite{ref20} and Riess {\it et al} \cite{ref21}
strongly favour a significant and positive $\Lambda$. Their finding arise from 
the study of more than $50$ type Ia supernovae with redshifts in the range
$0.10 \leq z \leq 0.83$ and suggest Friedmann models with negative pressure
matter such as a cosmological constant, domain walls or cosmic strings (Vilenkin
\cite{ref22}, Garnavich {\it et al} \cite{ref23}. The main conclusion of these works
is that the expansion of the universe is accelerating. 
\newline
\par
Several ans$\ddot{a}$tz have been proposed in which the $\Lambda$ term decays 
with time (see Refs. Gasperini \cite{ref24,ref25}, Freese {\it et al} \cite{ref26},
$\ddot{O}$zer and Taha \cite{ref12}, Peebles and Ratra \cite{ref27}, Chen and 
Hu \cite{ref28}, Abdussattar and Viswakarma \cite{ref29}, Gariel and Le Denmat
\cite{ref30}, Pradhan {\it et al} \cite{ref31} ). Of the special interest is the
ansatz $\Lambda \propto S^{-2}$ (where $S$ is the scale factor of the
Robertson-Walker metric) by Chen and Wu \cite{ref28}, which has been 
considered/modified by several authors ( Abdel-Rahaman \cite{ref32}, 
Carvalho {\it et al} \cite{ref12}, Waga \cite{ref33}, Silveira and Waga \cite{ref34},
Vishwakarma \cite{ref35}).
\newline
\par
In most treatments of cosmology, cosmic fluid is considered as perfect fluid.
However, bulk viscosity is expected to play an important role at certain stages
of expanding universe \cite{ref36}$^-$\cite{ref38}. It has been shown that bulk
viscosity leads to inflationary like solution \cite{ref39}, and acts like a negative
energy field in an expanding universe \cite{ref40}. A number of authors have discussed
cosmological solutions with bulk viscosity in various context \cite{ref41}$^-$\cite{ref44}.
\newline
\par
It has been shown by Berman \cite{ref2} and Berman \& Gomide \cite{ref4} that 
all the phases of the universe, i.e., radiation, inflation and pressure-free, 
may be considered as particular cases of the deceleration parameter 
$q$~ =~ constant type, as 
\[q = -\frac{S\ddot{S}}{\dot{S}^2},\]
where dots stand for time derivatives. We extend this definition to the 
Robertson-Walker cosmological models. In the past few decades there have been
numerous modification of general relativity in which gravitational constant
$G$ varies with time \cite{ref45}. Considering the principle of absolute quark
confinement, Der Sarkissian \cite{ref46} has suggested that gravitational and
cosmological constants may be considered as functions of time parameter in
Einstein's theory of relativity. A number of authors \cite{ref47}$^-$\cite{ref50}    
have considered time-varying $G$ and $\Lambda$ within the frame work of general 
relativity. 
\newline
\par
Maharaj and Naidoo \cite{ref51} utilised a form of the Hubble parameter 
to generate a number of solutions to the Einstein field equations with variable 
cosmological constant and variable gravitational constant in presence of a 
perfect fluid as the source of matter. These authors have extended the results 
obtained by Berman \cite{ref2,ref3}, Berman and Gomide \cite{ref4} 
by obtaining solutions to Einstein field equations, with variable gravitational 
and cosmological constants, in the Robertson-Walker spacetime. Explicit forms
for the gravitational constant, cosmological constant, scale factor, energy 
density and pressure are obtained for various cases.
\newline
\par
Motivated by the situations discussed above that bulk viscosity, gravitational and
cosmological ``constants'', are more relevant during early stages of the universe,
our intension in this paper is to extend the results obtained by Berman \cite{ref3},
Maharaj and Naidoo \cite{ref51} by including a bulk viscous fluid as a source of 
matter in the energy momentum tensor. This paper is organized as follows. In section 2, 
the solutions and main results of Berman \cite{ref3}, Maharaj and Naidoo \cite{ref51} 
are reviewed. In section 3, the corresponding solutions for a universe filled 
with a bulk viscous fluid are found. In section 4, a number of classes of new 
solutions for all cases of $k: 0, 1, -1$ for variable $G$ and $\Lambda$ are presented.
Our results are discussed in section 5.
\newline
\par

%%%%%%%%%%%%%%%%%%%%%%%%%%%%%%%%%%%%%%%%%%%%%%%%%%%%%%%%%%%%%%%%%%%%%%%%
%%%%%%%%%%%%%%%%%%%%    SECTION 2   %%%%%%%%%%%%%%%%%%%%%%%%%%%%%%%%%%%
%%%%%%%%%%%%%%%%%%%%%%%%%%%%%%%%%%%%%%%%%%%%%%%%%%%%%%%%%%%%%%%%%%%%%%%%

\section{Robertson-Walker Spacetime-Revisited}
In the standard coordinates $(x^a)$ = $(t, r, \theta, \phi)$ the Robertson-Walker
line element has the form
\begin{equation}
\label{eq1}
ds^{2} = -dt^{2} + S^{2}(t)\left[ \frac{dr^{2}}{1-kr^{2}} + r^{2} (d\theta^{2} + sin^{2}
\theta d\phi^{2})\right],
\end{equation}
\noindent 
where S(t) is a cosmic scale factor. Without loss of generality the constant
$k$ is related to the spatial geometry of a 3-dimensional manifold generated by $t$
= constant. The Robertson-Walker spacetimes are the standard cosmological models and
are consistent with observational results. For the case of variable cosmological
constant $\Lambda(t)$ and gravitational constant $G(t)$ the Einstein field equations
\begin{equation}
\label{eq2}
G_{ab} + \Lambda g_{ab} = 8\pi G T_{ab}
\end{equation}
\noindent 
yield
\begin{equation}
\label{eq3}
\frac{3}{S^2}(\dot{S}^2 + k) = 8\pi G\mu + \Lambda
\end{equation}
\begin{equation}
\label{eq4}
2\frac{\ddot{S}}{S} + \frac{(\dot{S}^2 + k)}{S^2} = -8\pi G p +\Lambda
\end{equation}
\noindent
for the line element (1). From Eqs. (\ref{eq3}) and (\ref{eq4}) we obtain the
generalised continuity equation
\begin{equation}
\label{eq5} 
\dot{\mu} + 3\frac{\dot{S}}{S}(\mu + p) + \frac{\dot{G}}{G} \mu + 
\frac{\dot{\Lambda}}{8\pi G} = 0.
\end{equation}

This reduces to the conventional continuity equation for constant  $\Lambda$ 
and $G$. If the classical conservation law, $T^{ab}_{;b} = 0$, also holds,
then Eq. (\ref{eq5}) implies two relationships
\begin{equation}
\label{eq6} 
\dot{\mu} + 3\frac{\dot{S}}{S}(\mu + p) = 0,
\end{equation}
\begin{equation}
\label{eq7} 
8\pi\mu\dot{G} + \dot{\Lambda} = 0,
\end{equation}
\noindent
which facilitate the solutions of the field equations. The result (\ref{eq6}) is just
the conventional continuity equation, and (\ref{eq7}) simply relates $G$ and $\Lambda$
and does not explicitly contain the scale factor $S(t)$. \\

Maharaj and Naidoo \cite{ref51} considered the generalised Einstein field equations 
(\ref{eq3})-(\ref{eq4})with variable gravitational constant $G$ and variable cosmological
constant $\Lambda(t)$ for the Robertson-Walker metric (1). They assumed the variation
of the Hubble parameter as given by the equation
\begin{equation}
\label{eq8} 
H \equiv  \frac{\dot{S}}{S} = D S^{-m}
\end{equation}
where $D$ and $m$ are constants. Then this imply that the deceleration parameter $q$
is constant i. e. $q = m -1$.

The form of the Hubble parameter (\ref{eq8}) was first utilised by Berman \cite{ref2}
and Berman and Gomide \cite{ref4} for the case of classical Einstein field equations with 
$\dot{\Lambda} = 0$ and $\dot{G} = 0$. Berman \cite{ref3} presented a solution to the
field equations (\ref{eq2}) for the $k =0$ Robertson-Walker spacetime:
\begin{equation}
\label{eq9} 
S = (C + m D t)^{1/m},
\end{equation}
\begin{equation}
\label{eq10} 
\Lambda = B S^{-2m},
\end{equation}
\begin{equation}
\label{eq11} 
G = \beta S^{mB/(4\pi A)},
\end{equation}
\begin{equation}
\label{eq12} 
\mu = \frac{A}{\beta} S^{-2m - m B /(4\pi A)},
\end{equation}
\begin{equation}
\label{eq13} 
p = \frac{A}{3\beta}\left [m(2 + \frac{B}{4\pi A}) - 3 \right ]S^{-2m - m B /(4\pi A)},
\end{equation}
where $A$, $B$, $C$, $\beta$ are constants and are subject to the following condition \\ 
\[3 D^2 = 8\pi A + B \] \\

It is worth noting that Eq. (\ref{eq16}) given by Berman \cite{ref3} corresponding
to Maharaj and Naidoo \cite{ref51} equation (\ref{eq13}), has an incorrect coefficient
on the right hand side. Equations (\ref{eq9})-(\ref{eq13}) comprise the general solution 
to the generalised Einstein field equations (\ref{eq3})-(\ref{eq4}) with variable
$G$ and $\Lambda$ for the Hubble law (\ref{eq8}). \\

%%%%%%%%%%%%%%%%%%%%%%%%%%%%%%%%%%%%%%%%%%%%%%%%%%%%%%%%%%%%%%%%%%%%%%%%%%%%
%%%%%%%%%%%%%%%%%%%%%  SECTION 3  %%%%%%%%%%%%%%%%%%%%%%%%%%%%%%%%%%%%%%%%%%
\section{Bulk Viscous Solutions of the Field Equations}
In this section bulk viscous models of the universe are discussed. 
Weinberge \cite{ref9} has suggested that in order to consider the 
effect of bulk viscosity, the perfect fluid pressure should be replaced 
by the effective pressure $\bar{p}$ by
\begin{equation}
\label{eq14}
\bar{p} = p - \xi \theta,
\end{equation}
where $\xi$ is the coefficient of bulk viscosity and $\theta$ is the 
expansion scalar given by $\theta = 3 H$. Here $\xi$ is, in general, a 
function of time. \\
Therefore, from Eq. (\ref{eq13}), we obtain
\begin{equation}
\label{eq15}
p - \xi \theta =  \frac{A}{3\beta} \left [m(2 + \frac{B}{4\pi A}) - 3 \right ]
S^{-2m - m B/(4\pi A)}
\end{equation}

For complete determinacy of the system, we assume an equation of state of an 
ideal gas given by
\begin{equation}
\label{eq16}
p = \gamma \mu,~ ~ ~ 0 \leq \gamma \leq 1,
\end{equation}
where $\gamma$ is a constant. Thus, given $\xi(t)$ we can solve for 
the cosmological parameters. It is standard to assume \cite{ref42,ref52} 
the following widely accepted {\it ad hoc} law
\begin{equation}
\label{eq17}
\xi(t) = \xi_0 \mu^n
\end{equation}
If $n = 1$, Eq. (\ref{eq17}) may correspond to a radiating fluid, 
whereas $n = 3/2$ may correspond to a string-dominated universe \cite{ref53}. 
However, more realistic models \cite{ref54} are based on $n$ lying the regime 
$0 \leq n \leq 1/2$. On using Eq. (\ref{eq17}) in Eq. (\ref{eq15}), we obtain
\begin{equation}
\label{eq18}
p - \xi_0 \mu^n \theta = \frac{A}{3\beta} \left [m(2 + \frac{B}{4\pi A}) - 3 \right ]
S^{-2m - m B/(4\pi A)}
\end{equation}

%%%%%%%%%%%%%%%%%%%%%%%%%%%%%%%%%%%%%%%%%%%%%%%%%%%%%%%%%%%%%%%%%%%%%%%%%
%%%%%%%%%%%%%%%%%%%%%%%   SUBSECTION 3.1  %%%%%%%%%%%%%%%%%%%%%%%%%%%%%%%

\subsection {Model I: $(\xi = \xi_0)$}
\noindent When $n = 0$, Eq. (\ref{eq17}) reduces to $\xi = \xi_0$ = constant 
and hence Eq. (\ref{eq18}) with Eq. (\ref{eq16}) gives
\begin{equation}
\label{eq19}
\mu = \frac{S^{-m}}{\gamma} \left[ 3 D \xi_0 + \frac{A}{3\beta} \left [m(2 + \frac{B}
{4\pi A}) - 3 \right ]\right]S^{-2m - m B/(4\pi A)}
\end{equation}

%%%%%%%%%%%%%%%%%%%%%%%%%%%%%%%%%%%%%%%%%%%%%%%%%%%%%%%%%%%%%%%%%%%%%%%%%%%%%%%%%%
%%%%%%%%%%%%%%%%%%%%%%%%%   SUBSECTION 3.2   %%%%%%%%%%%%%%%%%%%%%%%%%%%%%%%%%%%%%

\subsection {Model II: $(\xi = \xi_0 \mu)$}
\noindent When $n = 1$, Eq. (\ref{eq17}) reduces to $\xi = \xi_0 \mu$ and hence 
Eq. (\ref{eq18}) with  Eq. (\ref{eq16}) gives
\begin{equation}
\label{eq20}
\mu = \frac{A}{3 \beta(\gamma - 3D \xi_0 S^{-m})}  
\left [m(2 + \frac{B}{4\pi A}) - 3 \right ]
S^{-2m - m B/(4\pi A)}
\end{equation}
It is possible to avoid the horizon and monopole problem with the above variable $G(t)$
and $\Lambda(t)$ solutions as suggested by Berman \cite{ref3}. Other models are also 
considered by Berman {\it et al} \cite{ref55} and Bertolami \cite{ref47} which have 
the relationship
\[\Lambda \propto \frac{1}{t^2}. \]

This form of $\Lambda$ is physically reasonable as
observations suggest that $\Lambda$ is very small in the present universe. A decreasing
functional form permits $\Lambda$ to be large in the early universe. A partial list of
cosmological models in which the gravitational constant $G$ is decreasing function of
time are contained in Gr\o n \cite{ref41}, Hellings {\it et al} \cite{ref56},
Rowan-Robinson \cite{ref57}, Shapiro {\it et al} \cite{ref58} and Van Flandern \cite{ref59}.
The possibility of $G$ increasing with time, at least in some stages of the development 
of the universe, has been investigated by Abdel-Rahman \cite{ref32}, Chow \cite{ref60},
Levitt \cite{ref61} and Milne \cite{ref62}. In these models, by selecting proper signs
of different constants, one can make $\mu$ positive and decreasing function of time $t$.\\  

%%%%%%%%%%%%%%%%%%%%%%%%%%%%%%%%%%%%%%%%%%%%%%%%%%%%%%%%%%%%%%%%%%%%%%%%%%%%%%%
%%%%%%%%%%%%%%%%%%%%%%%%%%%%%%%%% SECTION 4  %%%%%%%%%%%%%%%%%%%%%%%%%%%%%%%%%%

\section{Other Solutions-Revisited}
\noindent Maharaj and Naidoo \cite{ref51} presented a number of classes of new solutions
for all classes of $k: 0, 1, -1$ for variable cosmological constant $\Lambda$ and
variable gravitational constant $G$ for Hubble law (\ref{eq8}). These solutions 
covered both the cases of $m = 0$ and $m\ne 0$ for the scale factor S:
\begin{equation}
\label{eq21}
S(t) = \left[ \begin{array}{ll}
              [C + m D t]^{1/m}  & \mbox{when $m\neq0$}\\
               Ee^{Dt}                        & \mbox{when $m=0$}

         \end{array} \right. 
\end{equation}
where C, D, E are constants. To solve the Einstein field equations (\ref{eq3})-(\ref{eq4})
they adopted the ansatz
\begin{equation}
\label{eq22}
\frac{3D^2}{S^{2m}} - \Lambda = K,
\end{equation}
\begin{equation}
\label{eq23}
8\pi G \mu - \frac{3k}{S^2} = K
\end{equation}
where $K$ is constant. This ansatz has the advantage of providing further classes of
solutions. Maharaj and Naidoo \cite{ref51} obtained the expressions for cosmological
constant $\Lambda$, energy density $\mu$ in terms of the gravitational constant $G$ 
and the scale factor $S$
\begin{equation}
\label{eq24}
\Lambda = \frac{3D^2}{S^{2m}} - K
\end{equation}
\begin{equation}
\label{eq25}
\mu = \frac{1}{8 \pi G}\left[\frac{3k}{S^2} + K\right]
\end{equation}

On substituting Eq. (\ref{eq25}) and the derivative of Eq. (\ref{eq24}) with respect to time
coordinate $t$ into Eq. (\ref{eq7}) we obtain the differential equation
\begin{equation}
\label{eq26} 
\frac{\dot{G}}{G} = 6 m D^2 \frac{\dot{S}}{S^{2m + 1}}\frac{1}{(\frac{3k}{S^2} + K)}
\end{equation}
relating $G$ to $S$. Thus the scalar factor $S$ is specified by our assumed form of the 
Hubble parameter, the gravitational constant $G$ is known in principle. The ansatz
(\ref{eq22})-(\ref{eq23}) enables to integrate all the Einstein field equations for a 
number of values of $m$, $k$ and $K$. In the remainder of this section we present a variety
of classes of solutions to the Einstein field equations for each of these cases considered 
in the presence of bulk viscosity. We list the form of the scale factor $S$, the variable 
cosmological constant $\Lambda$, the variable gravitational constant $G$, the energy
density $\mu$ and the effective pressure $\bar{p}$. There are other classes of solution
possible for other values of $m$. However the integration becomes extremely
complicated for general $m$, we present only some  simple cases in the following.\\

%%%%%%%%%%%%%%%%%%%%%%%%%%%%%%%%%%%%%%%%%%%%%%%%%%%%%%%%%%%%%%%%%%%%%%%%%%%%%%%%%%%%
%%%%%%%%%%%%%%%%%%%%%%%%%%%%%%%%%  SUBSECTION 4.1 %%%%%%%%%%%%%%%%%%%%%%%%%%%%%%%%%%

\subsection{Case I : $m = 0, K \ne 0, k \ne 0$.}
\noindent In this case we obtain
\begin{equation}
\label{eq27} 
S = E e^{Dt},
\end{equation}
\begin{equation}
\label{eq28} 
\Lambda = 3 D^2 - K,
\end{equation}
\begin{equation}
\label{eq29} 
G = A,
\end{equation}
\begin{equation}
\label{eq30} 
\mu = \frac{1}{8\pi A}\left[\frac{3k}{S^2} + K\right],
\end{equation}
\begin{equation}
\label{eq31} 
\bar{p} = - \frac{1}{8\pi A}\left[\frac{k}{S^2} + K\right].
\end{equation}

%%%%%%%%%%%%%%%%%%%%%%%%%%%%%%%%%%%%%%%%%%%%%%%%%%%%%%%%%%%%%%%%%%%%%%%%%%%%%%%%%%%%
%%%%%%%%%%%%%%%%%%%%%%%%%%% SUBSUBCASE 4.1.1  %%%%%%%%%%%%%%%%%%%%%%%%%%%%%%%%%%%%%%

\subsubsection{Model I : $(\xi = \xi_0)$}
\noindent When $ n=0 $, Eq. (\ref{eq17}) reduces to $\xi = \xi_0 $ = constant and hence
Eq. (\ref{eq31}) with Eq. (\ref{eq16}) gives
\begin{equation}
\label{eq32} 
\mu =  \frac{1}{\gamma} \left[ 3 \xi_0 D - \frac{1}{8\pi A}\left[\frac{k}{S^2}+K\right]\right].
\end{equation}

In this model, if we set $A < 0$ and $\xi_0, D, K, k > 0$ then $\mu$ is always positive
and decreasing function with time.\\  

%%%%%%%%%%%%%%%%%%%%%%%%%%%%%%%%%%%%%%%%%%%%%%%%%%%%%%%%%%%%%%%%%%%%%%%%%%%%%%%%%%%
%%%%%%%%%%%%%%%%%%%%%%%%%%% SUBSUBCASE 4.1.2  %%%%%%%%%%%%%%%%%%%%%%%%%%%%%%%%%%%%

\subsubsection{Model II : $(\xi = \xi_0\mu)$}
\noindent When $ n=1 $, Eq. (\ref{eq17}) reduces to $\xi = \xi_0\mu$ and hence
Eq.(\ref{eq31}) with Eq. (\ref{eq16} gives
\begin{equation}
\label{eq33} 
\mu = - \frac{1}{8\pi A (\gamma - 3 D \xi_0)} \left[\frac{k}{S^2} + K \right].
\end{equation}

In these de Sitter-type solutions $\Lambda$ and $G$ are strictly constant because of the
restriction $m = 0$. The cosmological constant $\Lambda$ vanishes when $K = 3D^2$ and
is positive for $K<3D^2$. The scale factor $S$ is exponential in $t$, so that if $D>0$
then the universe is exponentially expanding always. Such models are not physical
description of our present universe but could be applicable in the early universe in
the inflationary scenario. For $m =0$, we get the deceleration parameter $q = -1$
for these class of solutions which is consistent with the recent observations of 
supernovae Ia which require that the present universe is accelerating \cite{ref20,ref21}.
In model II, if we set $A < 0$, and $\xi_0, D, K, k > 0$  and $\gamma > 3D\xi_0$,
then $\mu$ is always positive and decreasing function with time.\\  

%%%%%%%%%%%%%%%%%%%%%%%%%%%%%%%%%%%%%%%%%%%%%%%%%%%%%%%%%%%%%%%%%%%%%%%%%%%%%%%%%%%%%%%%%%%%
%%%%%%%%%%%%%%%%%%%%%%%%   SUBSECTION 4.2   %%%%%%%%%%%%%%%%%%%%%%%%%%%%%%%%%%%%%%%%%%%%%%%%

\subsection{Case II : $m \ne 0, K = 0, k \ne 0$.} 
\noindent In this case we obtain
\begin{equation}
\label{eq34} 
S = [C + m D t]^{1/m},
\end{equation}
\begin{equation}
\label{eq35} 
\Lambda = \frac{3 D^2}{S^{2m}},
\end{equation}
\begin{equation}
\label{eq36} 
G = \alpha ~ exp\{{\frac{mD^2}{k(1 - m)} S^{2-2m}}\},
\end{equation}
\begin{equation}
\label{eq37} 
\mu = \frac{3k}{8 \pi \alpha} S^{-2}exp\{{\frac{mD^2}{k(1 - m)} S^{2-2m}}\},
\end{equation}
\begin{equation}
\label{eq38} 
\bar{p} = -\frac{1}{8 \pi \alpha} \left [\frac{4 m D^3}{S^{3m}}+\frac{3k}{S^2} \right ]
exp\{{\frac{mD^2}{k(1 - m)} S^{2-2m}}\}.
\end{equation}

%%%%%%%%%%%%%%%%%%%%%%%%%%%%%%%%%%%%%%%%%%%%%%%%%%%%%%%%%%%%%%%%%%%%%%%%%%%%%%%%%%%%%
%%%%%%%%%%%%%%%%%%%%%%%%   SUBSUBSECTION 4.2.1   %%%%%%%%%%%%%%%%%%%%%%%%%%%%%%%%%%%%

\subsubsection{Model I : $(\xi = \xi_0)$} 
\noindent When $n = 0$, Eq. (\ref{eq17}) reduces to $\xi = \xi_0 $ = constant and hence 
Eq. (\ref{eq38}) with Eq. (\ref{eq16}) gives
\begin{equation}
\label{eq39} 
\mu = \frac{1}{\gamma} \left[3 D \xi_0 S^{-m} -\frac{1}{8 \pi \alpha} \left [\frac{4 m D^3}
{S^{3m}}+\frac{3k}{S^2}\right]exp\{\frac{mD^2}{k(1 - m)} S^{2-2m}\}\right].
\end{equation}

%%%%%%%%%%%%%%%%%%%%%%%%%%%%%%%%%%%%%%%%%%%%%%%%%%%%%%%%%%%%%%%%%%%%%%%%%%%%%%%%%%%%%%%%
%%%%%%%%%%%%%%%%%%%%%%%%   SUBSUBSECTION 4.2.2   %%%%%%%%%%%%%%%%%%%%%%%%%%%%%%%%%%%%%%%

\subsubsection{Model II : $(\xi = \xi_0\mu)$} 
\noindent When $n = 1$, Eq. (\ref{eq17}) reduces to $\xi = \xi_0\mu$ and hence 
Eq.(\ref{eq38}) with Eq. (\ref{eq16}) gives
\begin{equation}
\label{eq40} 
\mu = - \frac{1}{8\pi \alpha (\gamma - 3 D \xi_0 S^{-m})} \left [\frac{4 m D^3}{S^{3m}}+
\frac{3k}{S^2} \right ]
exp\{{\frac{mD^2}{k(1 - m)} S^{2-2m}}\}.
\end{equation}
This case shares the common feature that $G$ may be increasing in time in certain regions
of spacetime with the model proposed by Abdel-Rahman \cite{ref32}.\\ 

%%%%%%%%%%%%%%%%%%%%%%%%%%%%%%%%%%%%%%%%%%%%%%%%%%%%%%%%%%%%%%%%%%%%%%%%%%%%%%%%%%%%%%%%
%%%%%%%%%%%%%%%%%%%%%%%     SUBSECTION 4.3  %%%%%%%%%%%%%%%%%%%%%%%%%%%%%%%%%%%%%%%%%%%%

\subsection{Case III : $m \ne 0, K \ne 0, k = 0$.}
\noindent In this case we obtain
\begin{equation}
\label{eq41} 
S = [C + m D t]^{1/m},
\end{equation}
\begin{equation}
\label{eq42} 
\Lambda = \frac{3 D^2}{S^{2m}} - K,
\end{equation}
\begin{equation}
\label{eq43} 
G = \alpha ~ ~ exp\{{\frac{3 D^2}{K S^{2m}}}\},
\end{equation}
\begin{equation}
\label{eq44} 
\mu = \frac{K}{8\pi \alpha}exp\{{-\frac{3D^2}{KS^{2m}}}\},
\end{equation}
\begin{equation}
\label{eq45} 
\bar{p} = -\frac{1}{8\pi \alpha}\left [\frac{2m D^2}{S^{2m}} + K \right ]
exp\{{-\frac{3D^2}{KS^{2m}}}\}.
\end{equation}

%%%%%%%%%%%%%%%%%%%%%%%%%%%%%%%%%%%%%%%%%%%%%%%%%%%%%%%%%%%%%%%%%%%%%%%%%%%%%%%%%%%%%%
%%%%%%%%%%%%%%%%%%%%%%%%   SUBSUBSECTION 4.3.1   %%%%%%%%%%%%%%%%%%%%%%%%%%%%%%%%%%%%%

\subsubsection{Model I : $(\xi = \xi_0)$} 
\noindent When $n = 0$, Eq. (\ref{eq17}) reduces to $\xi = \xi_0 $ = constant  and hence 
Eq. (\ref{eq45}) with Eq. (\ref{eq16}) gives
\begin{equation}
\label{eq46} 
\mu = \frac{1}{\gamma}\left[3 D \xi_0 S^{-m} -\frac{1}{8 \pi \alpha}
\left[\frac{2 m D^3}{S^{2m}}
+ K \right ] exp\{{-\frac{3D^2}{KS^{2m}}}\}\right].
\end{equation}

%%%%%%%%%%%%%%%%%%%%%%%%%%%%%%%%%%%%%%%%%%%%%%%%%%%%%%%%%%%%%%%%%%%%%%%%%%%%%%%%%%%%%%%
%%%%%%%%%%%%%%%%%%%%%%%%   SUBSUBSECTION 4.3.2   %%%%%%%%%%%%%%%%%%%%%%%%%%%%%%%%%%%%%

\subsubsection{Model II : $(\xi = \xi_{0} \mu)$} 
\noindent When n = 1, Eq. (\ref{eq17}) reduces to $\xi = \xi_{0} \mu$ and hence 
Eq. (\ref{eq45}) with Eq. (\ref{eq16}) gives
\begin{equation}
\label{eq47} 
\mu =  - \frac{1}{8\pi \alpha (\gamma - 3 D \xi_0 S^{-m})}\left [\frac{2m D^2}{S^{2m}}
+K\right ]exp\{{\frac{-3D^2}{KS^{2m}}}\}.
\end{equation}

%%%%%%%%%%%%%%%%%%%%%%%%%%%%%%%%%%%%%%%%%%%%%%%%%%%%%%%%%%%%%%%%%%%%%%%%%%%%%%%%%%%%%%%%
%%%%%%%%%%%%%%%%%%%%%%%%%%%%%  SUBSECTION 4.4  %%%%%%%%%%%%%%%%%%%%%%%%%%%%%%%%%%%%%%%%

\subsection{Case IV : $m =2, K \ne 0, k \ne 0$.}
\noindent In this case we obtain
\begin{equation}
\label{eq48} 
S = [C + 2 D t]^{1/2},
\end{equation}
\begin{equation}
\label{eq49} 
\Lambda = \frac{3 D^2}{S^4} - K,
\end{equation}
\begin{equation}
\label{eq50} 
G = \alpha \left[ \frac { (3kS^{-2} + K)^{K/k}} { exp \{S^{-2}\} } \right]^
{2D^{2}/k},
\end{equation}
\begin{equation}
\label{eq51} 
\mu = \frac{1}{8\pi \alpha}[ 3kS^{-2} + K] \left[\frac{exp \{S^{-2}\}}
{(3kS^{-2} + K)^{K/k}}\right]^{2D^{2}/k},
\end{equation}
\[
\bar{p} = - \frac{1}{8\pi \alpha}[3kS^{-2} + K] \left[\frac{exp \{S^{-2}\}}
{(3kS^{-2} + K)^{K/k}}\right]^{2D^{2}/k}\times 
\]
\begin{equation}
\label{eq52}   
\left[\frac{4D^2 S^{-4}(2K - 3k S^{-2})}{k(3 k S^{-2} + K)} + 1 \right]   
+\frac{kS^{-2}}{4\pi \alpha}\left[\frac{exp \{S^{-2}\}}{(3k S^{-2} + K)^
{K/k}}\right]^{2 D^{2}/k}.
\end{equation}

Unlike the cases considered thus far we have a specific value for $m$. 
This gives a value $q = 1$ for the deceleration parameter. A wide range 
of behaviour is possible for the gravitational constant.\\ 

%%%%%%%%%%%%%%%%%%%%%%%%%%%%%%%%%%%%%%%%%%%%%%%%%%%%%%%%%%%%%%%%%%%%%%%%%%%%%%%%%%%%%%%
%%%%%%%%%%%%%%%%%%%%%%%%%%%%%%%%%%  SUBSECTION 4.4.1 %%%%%%%%%%%%%%%%%%%%%%%%%%%%%%%%%

\subsubsection{Model I : $(\xi = \xi_{0})$}
\noindent When $n=0$, Eq. (\ref{eq17}) reduces to $\xi = \xi_{0}$ = constant and hence 
Eq. (\ref{eq52}) with Eq. (\ref{eq16}) gives
\[
\mu = \frac{3 D \xi_0 S^{-2}}{\gamma} - \frac{1}{8\pi \alpha \gamma}[3kS^{-2} + k] 
\left[\frac{exp \{S^{-2}\}}{(3kS^{-2} + K)^{K/k}}\right]^{2D^{2}/k}\times
\]
\begin{equation}  
\label{eq53}
\left[\frac{4D^2 S^{-4}(2K - 3k S^{-2})}{k(3 k S^{-2} + K)} + 1 - 
\frac{2kS^{-2}}{(3kS^{-2} + K)}\right].   
\end{equation}

%%%%%%%%%%%%%%%%%%%%%%%%%%%%%%%%%%%%%%%%%%%%%%%%%%%%%%%%%%%%%%%%%%%%%%%%%%%%%%%%%%%%%%%
%%%%%%%%%%%%%%%%%%%%%%%%%%%%%%%%%%  SUBSECTION 4.4.2 %%%%%%%%%%%%%%%%%%%%%%%%%%%%%%%%%%

\subsubsection{Model II : $(\xi = \xi_{0}\mu)$}
\noindent When $n=1$, Eq. (\ref{eq17}) reduces to $\xi = \xi_{0}\mu$ and hence 
Eq. (\ref{eq52}) with Eq. (\ref{eq16}) gives
\[
\mu = -\frac{1}{8\pi \alpha(\gamma - 3 D \xi_0 S^{-2})}[3kS^{-2} + K] 
\left[\frac{exp \{S^{-2}\}}{(3kS^{-2} + K)^{K/k}}\right]^{2D^{2}/k} \times
\]
\begin{equation} 
\label{eq54}
\left[\frac{4D^2 S^{-4}(2K - 3k S^{-2})}{k(3 k S^{-2} + K)} 
+ 1 - \frac{2kS^{-2}}{(3kS^{-2} + K)}\right].   
\end{equation}

%%%%%%%%%%%%%%%%%%%%%%%%%%%%%%%%%%%%%%%%%%%%%%%%%%%%%%%%%%%%%%%%%%%%%%%%%%%%%%%%%%%%%%%%
%%%%%%%%%%%%%%%%%%%%%%%%%%%%%  SUBSECTION 4.5 %%%%%%%%%%%%%%%%%%%%%%%%%%%%%%%%%%%%%%%%%

\subsection{Case V : $m = -2, K \ne 0, k \ne 0$.}
\noindent In this case we obtain
\begin{equation}
\label{eq55}
S = \frac{1}{\sqrt{C-2Dt}},
\end{equation}
\begin{equation}
\label{eq56}
\Lambda = \frac{3D^2}{S^{-4}} - K,
\end{equation}
\begin{equation}
\label{eq57}
G = \alpha \frac{exp\{\frac{3kS^2}{K^2} - \frac{S^4}{2K}\}}
{\left[S^{-2}(3kS^{-2} + K )\right]^{9k^{2}/K^{3}}},
\end{equation}
\begin{equation}
\label{eq58}
\mu = \frac{S^{-18k^{2}/K^{3}}}{8\pi \alpha}[3kS^{-2} + K]^{9k^{2}/K^{3} + 1} 
exp \{\frac{S^4}{2K} - \frac{3kS^2}{K^2}\},
\end{equation}
\[
\bar{p} = \frac{S^{-18k^{2}/K^{3}}}{4\pi \alpha}[3kS^{-2} + K]^{9k^{2}/K^{3} + 1}
\left[\frac{6k^2}{K^3} + \frac{3k - S^6}{3S^2(3kS^{-2} + K )} - \frac{1}{2} \right] \times 
\]
\begin{equation}
\label{eq59}
exp \{\frac{S^4}{2K}- \frac{3kS^2}{K^2}\}.
\end{equation}

%%%%%%%%%%%%%%%%%%%%%%%%%%%%%%%%%%%%%%%%%%%%%%%%%%%%%%%%%%%%%%%%%%%%%%%%%%%%%%%%%%%%%%
%%%%%%%%%%%%%%%%%%%%%%%%%%%%%%%%%%  SUBSUBSECTION 4.5.1 %%%%%%%%%%%%%%%%%%%%%%%%%%%%%%

\subsubsection{Model I : $(\xi = \xi_0)$}
\noindent When $n=0$, Eq. (\ref{eq17}) reduces to $\xi = \xi_0$ = constant and hence 
Eq. (\ref{eq59}) with Eq. (\ref{eq16}) gives
\[
\mu = \frac{3 D \xi_0 S^{2}}{\gamma} + \frac{S^{-18k^{2}/K^{3}}}{4\pi \alpha \gamma}
\left[3kS^{-2} + K\right]^{9k^{2}/K^{3} + 1}\times
\]
\begin{equation} 
\label{eq60}
\left[\frac{6k^2}{K^3} + \frac{3k - S^6}{3S^2(3kS^{-2} + K )} - \frac{1}{2} \right] 
exp \{\frac{S^4}{2K}- \frac{3kS^2}{K^2}\}.
\end{equation}

%%%%%%%%%%%%%%%%%%%%%%%%%%%%%%%%%%%%%%%%%%%%%%%%%%%%%%%%%%%%%%%%%%%%%%%%%%%%%%%%%%%%%%%
%%%%%%%%%%%%%%%%%%%%%%%%%%%%%%%%%%  SUBSUBSECTION 4.5.2 %%%%%%%%%%%%%%%%%%%%%%%%%%%%%%%

\subsubsection{Model II : $(\xi = \xi_0\mu)$}
\noindent When $n=1$, Eq (\ref{eq17}) reduces to $\xi = \xi_0\mu$ and hence Eq. (\ref{eq59})
with Eq. (\ref{eq16}) gives
\[
\label{eq61}
\mu = \frac{S^{-18k^{2}/K^{3}}}{4\pi \alpha(\gamma - 3 D \xi_0 S^{2})}[3kS^{-2} + K]^
{9k^{2}/K^{3} + 1}\times
\]
\begin{equation}
\left[\frac{6k^2}{K^3} + \frac{3k - S^6}{3S^2(3kS^{-2} 
+ K )} - \frac{1}{2}\right] exp \{\frac{S^4}{2K}- \frac{3kS^2}{K^2}\}.
\end{equation}
For $m = -2$, we get the deceleration parameter $q = -3$ for these class of solutions 
which is consistent with the recent observations of supernovae Ia which require the 
present universe is accelerating \cite{ref20,ref21} \\

%%%%%%%%%%%%%%%%%%%%%%%%%%%%%%%%%%%%%%%%%%%%%%%%%%%%%%%%%%%%%%%%%%%%%%%%%%%%%%%%%%%%%%%%
%%%%%%%%%%%%%%%%%%%%%%%%%%%%   SUBSECTION 4.6  %%%%%%%%%%%%%%%%%%%%%%%%%%%%%%%%%%%%%%%%%

\subsection{Case VI : $m = \frac{1}{2}, K \ne 0, k \ne 0$.}
\noindent In this case we obtain
\begin{equation}
\label{eq62}
S = [C + \frac{1}{2} D t]^2,
\end{equation}
\begin{equation}
\label{eq63}
\Lambda = \frac{3D^2}{S} - K,
\end{equation}
\begin{equation}
\label{eq64}
G = \alpha ~ exp \{\sqrt{\frac{3}{kK}} D^{2} arctan{(\sqrt{\frac{K}{3k}}S)}\},
\end{equation}
\begin{equation}
\label{eq65}
\mu = \frac{1}{8\pi \alpha}\left[3kS^{-2} + K\right] 
exp \{- \sqrt{\frac{3}{kK}} D^{2} arctan{(\sqrt{\frac{K}{3k}}S)}\},
\end{equation}
\begin{equation}
\label{eq66}
\bar{p} = - \frac{1}{8\pi \alpha}\left[(k - D^2 S) S^{-2} + K\right]
exp \{-\sqrt{\frac{3}{kK}}D^{2} arctan{(\sqrt{\frac{K}{3k}}S)}\}.
\end{equation}
For this class of solutions the deceleration parameter has the value 
$q=- \frac{1}{2}$ as $m = \frac{1}{2}$.\\

%%%%%%%%%%%%%%%%%%%%%%%%%%%%%%%%%%%%%%%%%%%%%%%%%%%%%%%%%%%%%%%%%%%%%%%%%%%%%%%%%%
%%%%%%%%%%%%%%%%%%%%%%%%%%%%%%%%%%  SUBSUBSECTION 4.6.1 %%%%%%%%%%%%%%%%%%%%%%%%%%

\subsubsection{Model I : $(\xi = \xi_0)$}
\noindent When $n=0$, Eq. (\ref{eq17}) reduces to $\xi = \xi_0$ = constant and hence 
Eq. (\ref{eq66})
with Eq. (\ref{eq16}) gives
\[
\mu = \frac{3 D \xi_0}{\gamma S^{1/2}}  - \frac{1}{8\pi \alpha}
\left[(k - D^2 S) S^{-2} + K\right]\times
\]
\begin{equation}
\label{eq67}
exp \{-\sqrt{\frac{3}{kK}} D^{2} arctan {(\sqrt{\frac{K}{3k}}S)}\}.
\end{equation}

%%%%%%%%%%%%%%%%%%%%%%%%%%%%%%%%%%%%%%%%%%%%%%%%%%%%%%%%%%%%%%%%%%%%%%%%%%%%%%%%%%%%
%%%%%%%%%%%%%%%%%%%%%%%%%%%%%%%%%%  SUBSUBSECTION 4.6.2 %%%%%%%%%%%%%%%%%%%%%%%%%%%%

\subsubsection{Model II : $(\xi = \xi_0\mu)$}
\noindent When $n=1$, Eq. (\ref{eq17}) reduces to $\xi = \xi_0\mu$ and hence 
Eq. (\ref{eq66})with Eq. (\ref{eq16}) gives
\[
\mu = - \frac{1}{8\pi \alpha(\gamma - 3D \xi_0 S^{ -1/2})}
\left[(k - D^{2} S) S^{-2} + K\right]\times
\]
\begin{equation}
\label{eq68}
exp \{-\sqrt{\frac{3}{kK}} D^{2} arctan {(\sqrt{\frac{K}{3k}}S)}\}.
\end{equation}
For these class of solutions, it is observed that  the universe is
accelerating.\\ 

%%%%%%%%%%%%%%%%%%%%%%%%%%%%%%%%%%%%%%%%%%%%%%%%%%%%%%%%%%%%%%%%%%%%%%%%%%%%%%%%%%%%%%
%%%%%%%%%%%%%%%%%%%%%%%%%%  SUBSECTION 4.7  %%%%%%%%%%%%%%%%%%%%%%%%%%%%%%%%%%%%%%%%%

\subsection{Case VII : $m = \frac{2}{3}, K \ne 0, k \ne 0$}
\noindent In this case we obtain
\begin{equation}
\label{eq69}
S = [C + \frac{2}{3} D t]^{3/2},
\end{equation}
\begin{equation}
\label{eq70}
\Lambda = \frac{3D^{2}}{S^{4/3}} - K,
\end{equation}
\begin{equation}
\label{eq71}
G = \alpha \left[\frac{(S^{2/3} + a )^{2}}{S^{4/3} -a S^{2/3} + a^{2}}
exp \{6 ~ arctan(\frac{2 S^{2/3} - a}{\sqrt{3} a})\} \right]^{D^{2}/K a^{2}},
\end{equation}
\begin{equation}
\label{eq72}
\mu = \frac{1}{8\pi \alpha}[3 k S^{-2} + K] \left[\frac{S^{4/3} - a S^{2/3} + a^{2}}
{(S^{2/3} + a)^{2}} exp \{-6 ~ arctan(\frac{2 S^{2/3} - a}{\sqrt{3} a})\} \right]
^{D^{2}/K a^{2}},
\end{equation}
\[
\bar{p} = - \frac{1}{8\pi \alpha} \left[\frac{S^{4/3} - a S^{2/3} + a^{2}}
{(S^{2/3} + a)^{2}}exp \{-6 ~ arctan(\frac{2 S^{2/3} - a}{\sqrt{3} a})\} \right]
^{D^{2}/K a^{2}}\times
\]
\[
\left [kS^{-2} + K + \frac{2D^2}{3Ka}(3k S^{-2} + K)\left[\frac{(S^{2/3} + a)^{2}}
{S^{4/3} - a S^{2/3} + a^{2}}\right] \right]\times
\]
\begin{equation}
\label{eq73}
\left[ \frac{S^{2/3} - a}{(S^{2/3} + a)^{3}} -
\frac{4\sqrt{3} (S^{4/3} - a S^{2/3} + a^{2})}{(S^{2/3} + a)^{2} \{3a^{2} + (2S^{2/3}
 - a)^{2}\}}\right].
\end{equation}
Here the deceleration parameter has the value $q = - \frac{1}{3}$ as $m = \frac{2}{3}$. 
For these class of solutions, it is observed that the universe is
accelerating.\\ 

%%%%%%%%%%%%%%%%%%%%%%%%%%%%%%%%%%%%%%%%%%%%%%%%%%%%%%%%%%%%%%%%%%%%%%%%%%%%%%%%%%%%%%
%%%%%%%%%%%%%%%%%%%%%%%%%%%%% SUBSUBSECTION 4.7.1      %%%%%%%%%%%%%%%%%%%%%%%%%%%%%%

\subsubsection{Model I : $(\xi = \xi_0)$}
\noindent When $n = 0$, Eq. (\ref{eq17}) reduces to $\xi = \xi_0$ = constant  
and hence Eq. (\ref{eq73}) with Eq. (\ref{eq16}) gives
\[
\mu = \frac{3D\xi_0}{\gamma S^{2/3}} - \frac{1}{8\pi \alpha \gamma} \left[\frac{S^{4/3} - 
a S^{2/3} + a^{2}}{(S^{2/3} + a)^{2}}exp \{-6 ~ arctan(\frac{2 S^{2/3} - a}{\sqrt{3} a})\}
 \right]^{D^{2}/K a^{2}}\times
\]
\[
\left[kS^{-2} + K + \frac{2D^2}{3Ka}(3k S^{-2} + K)\left[\frac{(S^{2/3} + a)^{2}}
{S^{4/3} - a S^{2/3} + a^{2}}\right] \right]\times
\].
\begin{equation}
\label{eq74}
\left[ \frac{S^{2/3} - a}{(S^{2/3} + a)^{3}} -
\frac{4\sqrt{3} (S^{4/3} - a S^{2/3} +a^{2})}{(S^{2/3} + a)^{2} \{3a^{2} + (2S^{2/3}
 - a)^{2}\}}\right].
\end{equation}

%%%%%%%%%%%%%%%%%%%%%%%%%%%%%%%%%%%%%%%%%%%%%%%%%%%%%%%%%%%%%%%%%%%%%%%%%%%%%%%%%%%%%%
%%%%%%%%%%%%%%%%%%%%%%%%%%%%% SUBSUBSECTION 4.7.2      %%%%%%%%%%%%%%%%%%%%%%%%%%%%%%%

\subsubsection{Model II : $(\xi = \xi_0\mu)$}
\noindent When $n = 0$, Eq. \ref{eq17}) reduces to $\xi = \xi_0\mu$  and hence 
Eq. (\ref{eq73}) with Eq. (\ref{eq16}) gives
\[
\mu = - \frac{1}{8\pi \alpha(\gamma - 3 D \xi_0 S^{-2/3})} \left[\frac{S^{4/3} - a S^{2/3} 
+ a^{2}}{(S^{2/3} + a)^{2}} exp \{-6 ~ arctan(\frac{2 S^{2/3} - a}{\sqrt{3} a})\} \right]
^{D^{2}/Ka^{2}}\times
\]
\[
\left[k S^{-2} + K + \frac{2D^2}{3Ka}(3k S^{-2} + K)
\left[\frac{(S^{2/3} + a)^{2}}{S^{4/3} - a S^{2/3} + a^{2}}\right] \right]\times
\]
\begin{equation}
\label{eq75}
\left[ \frac{S^{2/3} - a}{(S^{2/3} + a)^{2}} -
\frac{4\sqrt{3} (S^{4/3} - a S^{2/3} +a^{2})}{(S^{2/3} + a)^{2} \{3a^2 + (2S^{2/3}
 - a)^{2}\}}\right].
\end{equation}

In the above we have presented a number of new solutions to the Einstein field equations 
with variable cosmological constant and gravitational constant which satisfy the Hubble 
variation law given by Eq. (\ref{eq8}). It remarkable that this simple law leads to a 
wide class of solutions. It is interesting to observe that solutions are admitted in 
which the gravitational constant may be increasing with time (cf. Abdel-Rahman, 1990).
The anstaz utilised to solve the Einstein field equations (3)-(4) is very simple. It
might be worthwhile to investigate other possibilities that lead to solutions to the
Einstein's field equations with interesting behaviour of the gravitational constant
and cosmological constant. \\

\section {Conclusions}
In this paper we have investigated Einstein's equation in the presence of
a viscous fluid, for the Robertson-Walker universe within the framework
of general relativity, where the gravitational constant $G$ and the cosmological 
parameter $\Lambda$ are variables. We utilize a form for the Hubble parameter
($H = D S^{-m}$) to generate a number of solutions to Einstein field equations.
For these class of solutions where $m=0, -2, \frac{1}{2}, \frac{2}{3}$, we find 
the decelerating parameters as negative. These class of solutions are consistent
with the recent observations of supernovae Ia \cite{ref20,ref21} which require 
the present universe is accelerating. From our results we observe that $\mu$
is decreasing with time with suitable choice of constants whereas $G$ is increasing 
function of time. The possibility of an increasing $G$ has been suggested by 
Abdel-Rahaman \cite{ref32}, Arbab \cite{ref63} and Massa \cite{ref64}. \\

Assuming an {\it ad hoc} law of the form $\xi(t) = \xi_0 \mu^{n}$, where $\mu$
is the energy density and $n$ is the positive index, we have obtained exact 
solutions. The models discussed here are isotropic and homogeneous and, in
view of the assumption of isotropy the sheer viscosity is absent. The effect
of the bulk viscosity is to produce a change in the perfect fluid. We observe
that Murphy's conclusion \cite{ref53} about the absence of big bang type 
singularity in the finite past in models with bulk viscous fluid is, 
in general, not true.\\   
\nonumsection{Acknowledgements}
The authors would like to thank the Intre-University Centre for Astronomy and 
Astrophysics, Pune, India for hospitality where this work was carried out. 
We would lso like to thank R. G. Vishwakarma for his careful reading of the 
manuscript and the many useful suggestions.
\newline
\newline
%% \newpage
\nonumsection{References}

\end{document}